\documentclass[iop]{emulateapj}
\newcommand{\beq}{\begin{equation}}
\newcommand{\eeq}{\end{equation}}

\newcommand{\hi}{H{\sc i}~}

\newcommand{\hia}{H{\sc i}}

\newcommand{\citei}[1]{\citeauthor{#1} \citeyear{#1}}

\newcommand{\citeia}[2]{\citeauthor{#1}~(\citeyear{#1};~#2)}
\usepackage{color}
\usepackage{amsmath}
\slugcomment{}

\begin{document}

\title{A Comparison of Far IR and HI as Reddening Predictors at High Galactic Latitude}

\author{J.~E.~G.~Peek\altaffilmark{1}\altaffilmark{2}}
\altaffiltext{1}{Department of Astronomy, Columbia University, New York, NY. jegpeek@gmail.com}
\altaffiltext{2}{Hubble Fellow}

\begin{abstract}
Both the Galactic 21-cm line flux from neutral hydrogen (\hia) in interstellar medium and the far-infrared (FIR) emission from Galactic dust grains have been used to estimate the strength of Galactic reddening of distant sources. In this work we use a collection of uniform color distant galaxies as color standards to determine whether the \hi method or the FIR method is superior. We find that the two methods both produce reasonably accurate maps, but that both show significant errors as compared to the typical color of the background galaxies. We find that a mixture of the FIR and \hi maps in roughly a 2-to-1 ratio is clearly superior to either map alone. We recommend that future reddening maps should use both sets of data, and that well-constructed FIR and \hi maps should both be vigorously pursued. 
\end{abstract}


\section{Introduction}\label{intro}

The interstellar medium (ISM) is a very sooty material by the standards of Earth's atmosphere, suffused with obscuring astrophysical dust. Through emission, absorption, and scattering, this dust acts to modify the light we receive from distant objects across almost all observable wavelengths, most notoriously acting to artificially redden objects in the optical. One way we combat this distortion of our observations is to construct maps of the surface density of dust on the sky, such that observations can be corrected back to an unreddened state. As we push forward into an era of precision cosmology, it is especially important to have high accuracy maps at high Galactic latitude, where the bulk of cosmological investigations take place (e.~g.~ Huterer, Cunha, \& Fang, 2013, \citei{MKS10}).

In the past two main methods have been used to construct such reddening maps. The first is the \hi method. The 21-cm hyperfine transition of the hydrogen atom traces neutral gas throughout the Galactic ISM. At high latitudes the \hi ISM is largely optically thin (i.e. $\tau_{HI} < 1$; with some exceptions e.~g.~\citei{2011ApJ...735..129P}) such that the integrated 21-cm profile is a good proxy for the column density. Since the high Galactic latitude ISM has a largely uniform dust-to-gas ratio, and neutral gas dominates the column density, the 21-cm intensity can act as a proxy for dust column. This method was largely codified by \citet{Burstein:1978bk}, and maps were presented in \citeia{Burstein:1982dz}{BH}. A more modern database of \hi columns is presented in \citeia{Kalberla05}{LAB}, which has the advantages of being more coherently reduced, all sky, and having careful corrections for stray radiation, which can be a major source of error in \hi column density maps.

The \hi method acted as the standard for high Galactic latitude extinction until the far infrared (FIR) method was developed by \citeia{SFD98}{SFD}. In the FIR method the assumption is made that there is a single population of dust, both in terms of grain size distribution and composition, and that therefore there is a direct relationship between the FIR thermal emission from grains and their contribution to extinction and reddening. This relationship is subject to the variations in the temperature of the grains. To build an extinction map SFD made an all-sky map from the IRAS 100 $\mu$m data, as well as a dust temperature map from DIRBE. Combining these two, they presented an all-sky extinction map which has been the standard since its publication. The SFD map superseded the BH map due largely to its higher resolution, fidelity, completeness, and ease of use, rather than some known superiority of the FIR method over the \hi method. 

Both the \hi and FIR methods have inherent flaws that stem from their assumptions. In reality, the \hi column is not entirely optically thin, and thus the integrated 21-cm flux is not a perfect proxy for neutral column density. Furthermore, other phases of the ISM carry obscuring dust, including molecular and ionized phases. 
We also know that the dust-to-gas ratio is not a fixed quantity, even with these assumptions taken into account \citep{Burstein:1978bk}. There are also a number of known issues with the FIR method. SFD note that variation of grain temperature along the line of sight can generate significant errors. Grain size variation can also effect the results since FIR dust emission is biased toward larger grains, but extinction is biased toward smaller grains. Indeed, some authors have found dust emissivity is $\sim 3$ times higher in molecular gas than in atomic gas due to grain agglomeration (\citei{2011A&A...536A..19P}, \citei{Stepnik:2003iq}). There are also other sources of FIR emission, especially unresolved, distant galaxies, which can contaminate these maps (\citei{Yahata07}, \citei{Kashiwagi:2012we}).

The modern era of digitized optical surveys has allowed us to test the accuracy of these maps in some detail, specifically, with the extremely consistent photometric calibration of the \emph{Sloan Digital Sky Survey} (SDSS; \citei{york00}, \citei{Padmanabhan:2008js}). There has been a burst of activity using ``standard crayon'' methods; selections of large sets of objects corrected to have identical intrinsic colors such that variations from these colors must be due to effects from the intervening dust. \citet{Schlafly10} used photometric observations of large groups of stars, \citeia{2010ApJ...719..415P}{PG10} used spectroscopic observations of non-starforming galaxies, \citet{Jones:2011tf} used spectroscopic observations of cool stars, and \citeia{2011ApJ...737..103S}{SF11} used spectroscopic observations of a broader range of stars. All of these explorations found that SFD is largely accurate in the optical, but that overall errors and spatial variability of the reddening are detectable.

In this work we attempt to resolve whether the FIR method, the \hi method, or some combination make the most consistent reddening maps. To do this we employ the standard crayons developed in PG10 and compare maps of the residual colors of these galaxies to the FIR maps of SFD to the \hi column densities derived from LAB. To put the \hi and FIR methods on equal footing we work at a resolution lower than that of either survey, which also serves to aggregate enough PG10 galaxies to give statistically significant color variations from the expectations of SFD. In \S \ref{dm} we briefly summarize the methods used to produce the low-resolution maps of Galaxy colors, \hi, and FIR. In \S \ref{results} we show the relative accuracy of the two methods, and a combination of the methods. We discuss these results and conclude in \S \ref{discon}.

\section{Data and Methods}\label{dm}

Here we aim to generate three maps to compare the \hi method predictions and FIR method predictions to the observed color variation of galaxy colors on the sky. We would prefer to have \hi maps at a resolution comparable to the SFD FIR maps, but the only such large-area maps \citep{2011ApJS..194...20P} are neither complete nor corrected for stray radiation, and thus are at present unsuitable for such an analysis. Furthermore, to get to a high enough signal-to-noise to detect significant color variations from predictions we must average the PG10 galaxies over larger areas than even the LAB survey beam. We note that the higher resolution GASS survey \citep{McClureGriffiths:2009ui} would be a useful map for this analysis, were it in the same hemisphere as the SDSS data, and we expect the EBHIS maps will also be useful once complete and corrected for stray radiation \citep{Winkel10}. Given this state of affairs we employ the LAB survey column densities as our data set from which to generate our \hi method maps, and SFD for the FIR method maps. We note that while a 14\% correction to SFD was suggested by \citet{2011ApJ...737..103S}, it is not clear that this correction applies to the very low extinctions we are investigating at high latitude, and \citeia{Peek:2013vg}{PS13} find no evidence for such a variation in either the optical or UV. We therefore use the SFD reddening predictions with no correction. Throughout this text we will refer to the $E\left( B-V\right)$ reddening predictions from SFD as $E_{SFD}$ and the predictions from LAB as $E_{LAB}$. 


The 151,637 PG10 galaxies were selected from the SDSS DR7 main galaxy sample data release to have no detected OII or H$\alpha$, and thus be almost entirely non-starforming, passively evolving galaxies. The galaxies were K-corrected, corrected for an absolute magnitude-color relationship as a function of redshift, as well as corrected for very weak dependence of galaxy color on galaxy environmental density. The final variation in $g-r$ galaxy color translates to a typical standard deviation in $E\left( B-V\right)$ of 0.024 magnitudes. The details of this analysis are available in PG10. As in PS13, we employ the galaxy colors as uncorrected by SFD. 

We follow the same map-making methods as described in PS13. Essentially, we put the galaxies into a zenith equal area projection and grid them into 16 square degree pixels. To evaluate the $E\left( B-V\right)$ in these pixels we use the median color of the galaxies within each pixel, as the color distribution has outliers that can corrupt the average color. We refer to this value as $E_{PG10}$. The error in each pixel of the PG10 map is evaluated using a standard boostrap method. Each pixel has a typical noise of 2 millimagnitudes in measured $E\left( B-V\right)$. We restrict our investigation to areas of sky where we have errors in this map below 4 millimagnitudes. To provide a reliable comparison, we evaluate the SFD and LAB maps at the same positions as the galaxies within a given pixel and take the median value within the pixel. While in practice this makes little difference on our final results, it nicely preserves the weighting within each pixel that comes from the precise galaxy positions within the survey. This is especially relevant towards the edges of the survey. 

\section{Results}\label{results}

\begin{figure*}
\includegraphics[scale=1.0]{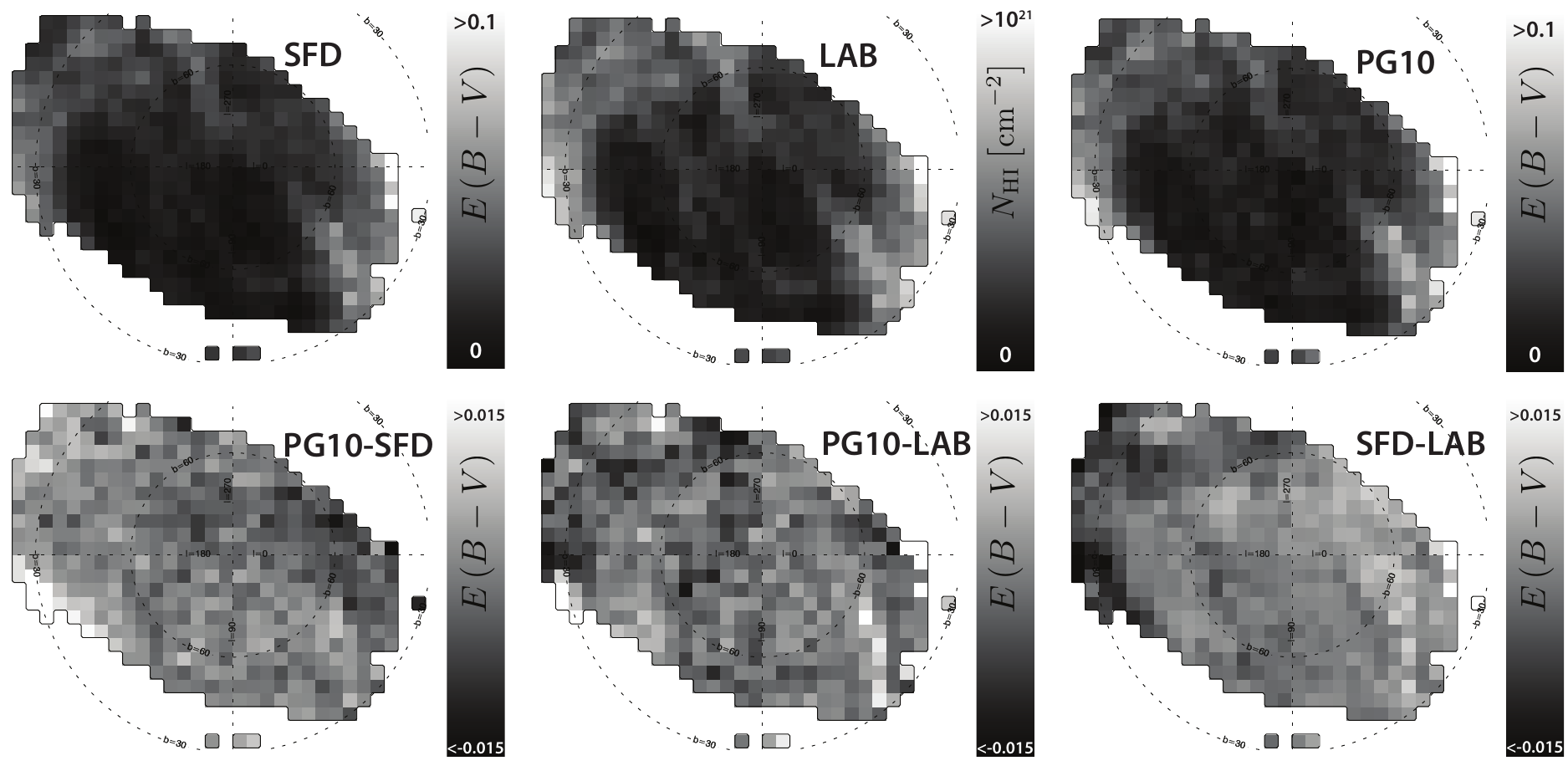}
\caption{Maps of reddening towards the northern Galactic cap. The top row shows the SFD, LAB, and PG10 maps described in \S \ref{dm}. We do not show the small fraction of sky covered toward the southern Galactic cap. The bottom row shows the difference between various maps. To take the difference we simply take the residual of a first-order polynomial error-weighted least-squares fit of the PG10 map to the SFD map and the PG10 data to the LAB map. The difference of these two fits is shown at the far right. We note that the color residuals from the \hi and LAB maps are both highly statisitically significant (with typical erorrs of 0.002) and are structurally different, implying that their errors stem from different effects in the ISM.}
\label{sixmaps}
\end{figure*}

The resulting maps of $E_{SFD}$, $N_{HI,LAB}$, and $E_{PG10}$ are shown in Figure \ref{sixmaps}. Also shown are the residuals when taking the difference between pairs of maps. To do this we perform a first-order polynomial error-weighted least-squares fit solving
\beq
E_{PG10} = a_0 + a_{SFD} E_{SFD}
\eeq
and 
\beq
E_{PG10} = b_0 + b_{LAB} N_{HI,LAB}.
\eeq
We find  $a_0 = -1.36 \times 10^{-3}, a_{SFD} = 1.10$ and $b_0 = -3.78 \times 10^{-3} , b_{LAB} = 1.44 \times 10^{-22}$. As this analysis is performed over a somewhat restricted area of sky we do not specifically recommend these values for future work, but rather report them to show broad consistency with previous work (e.~g.~ \citei{Burstein:1978bk}, \citep{2011ApJ...737..103S}). As expected, the three $E_{SFD}$, $N_{HI,LAB}$, and $E_{PG10}$ maps look rather similar. We note the that the residual maps (PG10-SFD, PG10-LAB) have many pixels that are significantly different than the expectation, in a statisitical sense. What is perhaps most interesting is that the PG10-SFD and PG10-LAB residual maps have inconsistent deviations from zero. This is to say, whatever processes make the \hi method fail are not the same processes that make the FIR method fail. 

\begin{figure}
\includegraphics[scale=1.0]{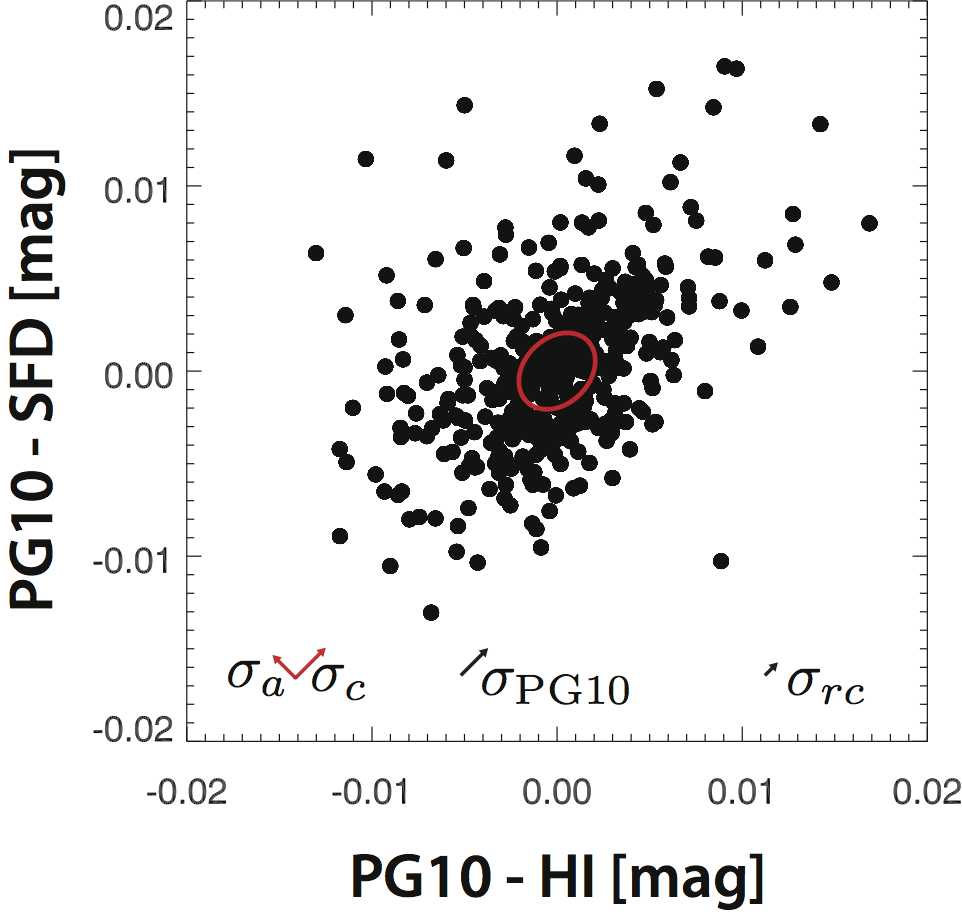}
\caption{A scatter plot of the PG10-SFD versus PG10-LAB maps, with filled circles representing each pixel. In red is the 1-$\sigma$ ellispe that shows the correlation (towards $y+x$) and anticorrelation (towards $y-x$) between the two maps. The amplitudes of these dispersions is shown at the lower left of the figure. At the bottom center of the figure is the expected correlation between the maps that stems simply from the known errors in PG10, $\sigma_{\rm PG10}$. At bottom right is the residual correlation between the maps, which we note is significantly smaller than the anticorrelation.}
\label{err_corr}
\end{figure}

To further quantify this discrepancy we analyze the map data in the PG10-SFD versus PG10-LAB space, as in Figure \ref{err_corr}. We find the residual maps are correlated with amplitude of $\sigma_c = 0.0023$ and the anti-correlated with amplitude $\sigma_a =0.0017$. The preponderance of the correlation stems from the noise in the PG10 data itself (which contributes to both maps) with mean $\sigma_{\rm PG10} =0.0020$ (see \S \ref{dm}). We find that the amplitude of the correlation in the residual maps that stem from the HI and FIR maps themselves is only $\sigma_{cr} =0.0010$, less than 60\% of the anticorrelation. We note that some fair fraction of $\sigma_{cr}$ may in fact be due to errors in the SDSS data or the PG10 sample which are not captured by our bootstrap error analysis. Imperfections in the SDSS photometric calibrations, for instance, would contribute to this error (see e.~g.~ PG10 figure 7). Even without appealing to sources of error beyond the statisitcal error in the PG10 map, the anticorrelation between the two residual maps is larger than the correlation, indicating that the FIR and HI maps are complementary.

\begin{figure}
\includegraphics[scale=1.0]{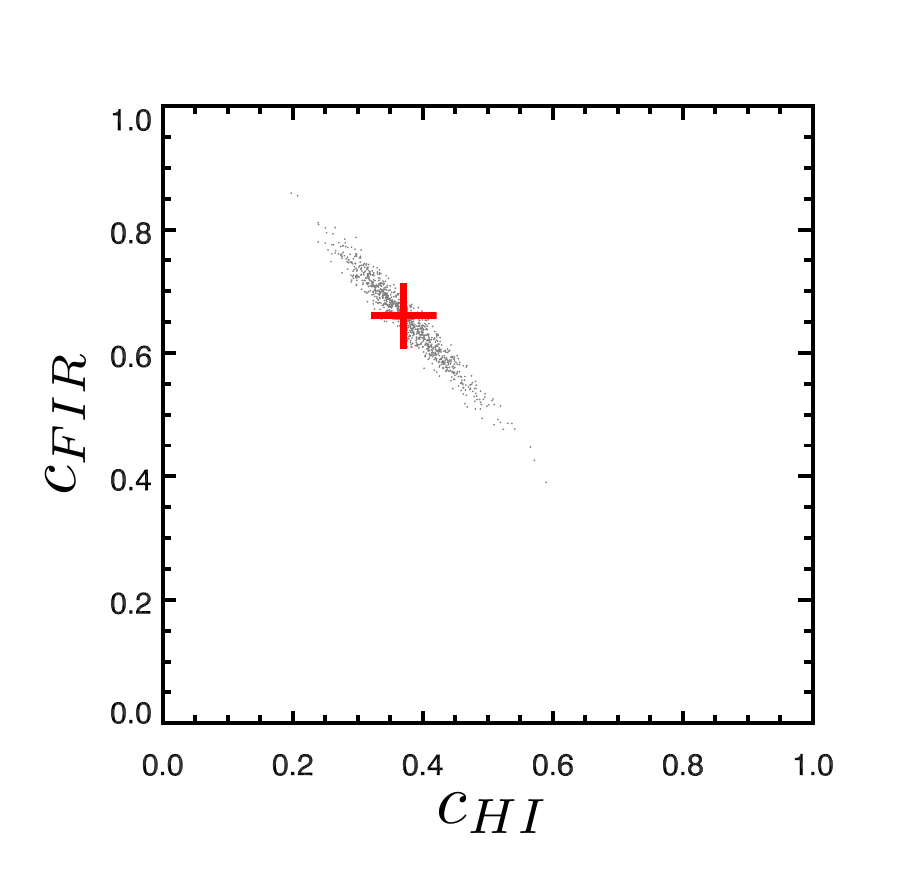}
\caption{The best fit to Equation \ref{ceqn} is shown as a red cross. Gray points are the results of 1000 Monte Carlo bootstraps, to give a sense of the errors. Clearly, a 66\% FIR + 37\% HI map is superior to either a 100\% FIR or 100\% \hi map.}
\label{HI_dep}
\end{figure}

With this in mind we perform a similar error-weighted least-squares fit to the following equation:
\beq\label{ceqn}
E_{PG10} = c_0 + c_{FIR} a_{SFD} E_{SFD} + c_{HI} b_{LAB} N_{HI,LAB}
\eeq
using the previously determined values of $a_{SFD}$ and $b_{LAB}$, and solving for $c_0, c_{FIR},$ and $c_{HI}$. Additionally, we run a simple bootstrap analysis to probe the error space, resampling the map pixels with replacement. The results are shown in Figure \ref{HI_dep}. We find best fit values of $c_{FIR}= 0.66, c_{HI} = 0.37$. This to say a healthy mixture of the two methods, with a weighting toward the FIR method, produces a significantly superior map to using either map alone. Errors in the PG10-SFD and PG10-LAB maps are 0.0042 magnitudes in both cases, as assesed using the standard deviation of the data clipped at 3.5 sigma. The equivalent error in the best fit mixture map is 0.0038 magnitudes.

\section{Discussion \& Conclusion}\label{discon}
Our analysis shows that while both the FIR method (in this case SFD) and the \hi method (in this case LAB) produce maps that are largely error-free, each map does have detectable variation from the observed reddening (PG10). Furthermore, we show that these variations are not well-correlated with one another -- whatever mechanism produces discrepancies in the \hi reddening map is not the same mechanism that produces discrepancies in the FIR reddening map. Quantitatively, an admixture of 66\% FIR and 37\% HI outperforms either map alone. 

This has significant implications for the construction of future reddening maps. Specifically, it shows that the FIR method and \hi method have complementary strengths, and thus to make the best possible map accurate, all-sky, high-resolution maps in both the FIR and the \hi should be constructed. Higher resolution FIR maps are already underway from the all-sky WISE and AKARI data (Finkbeiner, private communication), which will certainly feed into such a map. This result also has bearing on upcoming \hi surveys; the combined EBHIS + GASS map at 3 to 4 times the resolution of LAB, as well as much higher resolution maps to be completed by the WHNHS survey on APERTIF \citep{Verheijen:2009ua} and the WALLABY survey on ASKAP \citep{Duffy:2012ie}, both SKA precursor instruments. The careful construction of these data products should be considered a priority for these instruments, as their results will have significant bearing on much of the rest of observational astronomy. 

We note two caveats to this work. Firstly, our maps are constructed on a much larger angular scale than useful modern reddening maps. Thus, there is a possibility that one or another of these methods would become more dominant on smaller scales. We find this possibility unlikely, as it is clear that there are detectable large-scale errors in each map, which we expect to remain despite improvements in resolution. The second is that each one of these methods may be individually improved, thus rendering the other less useful. The FIR methods could be improved with higher fidelity and resolution temperature maps, as well as corrections for extragalactic sources of FIR light. The \hi maps could be improved with some ad hoc correction for \hi opacity, and perhaps the addition of molecular tracers like CO or OH, or even corrections for ionized gas. While these kinds of advancements are likely to improve the quality of each method, their basic weaknesses as described in \S \ref{intro} are unlikely to disappear, and so the combination of the two methods will still be superior to either one alone.

JEGP would like to thank Doug Finkbeiner, Eddie Schlafly, David Schiminovich, Karin Sandstrom, and Robert Goldston for helpful conversations. JEGP was supported by HST-HF-51295.01A, provided by NASA through a Hubble Fellowship grant from STScI, which is operated by AURA under NASA contract NAS5-26555. 


\begin{thebibliography}{22}
\expandafter\ifx\csname natexlab\endcsname\relax\def\natexlab#1{#1}\fi

\bibitem[{Ade {et~al.}(2011) Collaboration, Ade, Aghanim, Arnaud,
  Ashdown, Aumont, Baccigalupi, Balbi, Banday, Barreiro, Bartlett, Battaner,
  Benabed, Benoit, Bernard, Bersanelli, Bhatia, Bock, Bonaldi, Bond, Borrill,
  Bouchet, Boulanger, Bucher, Burigana, Cabella, Cardoso, Catalano, Cayon,
  Challinor, Chamballu, Chiang, Chiang, Christensen, Clements, Colombi,
  Couchot, Coulais, Crill, Cuttaia, Dame, Danese, Davies, Davis, de~Bernardis,
  de~Gasperis, de~Rosa, de~Zotti, Delabrouille, Delouis, Desert, Dickinson,
  Dobashi, Donzelli, Dore, Dorl, Douspis, Dupac, Efstathiou, Ensslin, Eriksen,
  Falgarone, Finelli, Forni, Fosalba, Frailis, Franceschi, Fukui, Galeotta,
  Ganga, Giard, Giardino, Giraud-Heraud, Gonzalez-Nuevo, Gorski, Gratton,
  Gregorio, Grenier, Gruppuso, Hansen, Harrison, Helou, Henrot-Versille,
  Herranz, Hildebrandt, Hivon, Hobson, Holmes, Hovest, Hoyland, Huffenberger,
  Jaffe, Jones, Juvela, Kawamura, Keihanen, Keskitalo, Kisner, Kneissl, Knox,
  Kurki-Suonio, Lagache, Lamarre, Lasenby, Laureijs, Lawrence, Leach, Leonardi,
  Leroy, Lilje, Linden-Vornle, Lopez-Caniego, Lubin, Macias-Perez, MacTavish,
  Maffei, Maino, Mandolesi, Mann, Maris, Martin, Martinez-Gonzalez, Masi,
  Matarrese, Matthai, Mazzotta, McGehee, Meinhold, Melchiorri, Mendes,
  Mennella, Miville-Desch{\^e}nes, Moneti, Montier, Morgante, Mortlock, Munshi,
  Murphy, Naselsky, Natoli, Netterfield, Norgaard-Nielsen, Noviello, Novikov,
  Novikov, O'Dwyer, Onishi, Osborne, Pajot, Paladini, Paradis, Pasian,
  Patanchon, Perdereau, Perotto, Perrotta, Piacentini, Piat, Plaszczynski,
  Pointecouteau, Polenta, Ponthieu, Poutanen, Prezeau, Prunet, Puget, Reach,
  Reinecke, Renault, Ricciardi, Riller, Ristorcelli, Rocha, Rosset,
  Rowan-Robinson, Rubino-Martin, Rusholme, Sandri, Santos, Savini, Scott,
  Seiffert, Shellard, Smoot, Starck, Stivoli, Stolyarov, Stompor, Sudiwala,
  Sygnet, Tauber, Terenzi, Toffolatti, Tomasi, Torre, Tristram, Tuovinen,
  Umana, Valenziano, Vielva, Villa, Vittorio, Wade, Wandelt, Wilkinson, Yvon,
  Zacchei, \& Zonca}]{2011A&A...536A..19P}
 Ade, P. A. et al. 2011, A\&A, 536, 19

\bibitem[{Burstein \& Heiles(1978)}]{Burstein:1978bk}
Burstein, D. \& Heiles, C. 1978, ApJ, 225, 40

\bibitem[{Burstein \& Heiles(1982)}]{Burstein:1982dz}
---. 1982, The ApJS, 87, 1165

\bibitem[{Duffy {et~al.}(2012)Duffy, Meyer, Staveley-Smith, Bernyk, Croton,
  Koribalski, Gerstmann, \& Westerlund}]{Duffy:2012ie}
Duffy, A.~R., Meyer, M.~J., Staveley-Smith, L., Bernyk, M., Croton, D.~J.,
  Koribalski, B.~S., Gerstmann, D., \& Westerlund, S. 2012, Monthly Notices of
  the Royal Astronomical Society, 426, 3385

\bibitem[{Jones {et~al.}(2011)Jones, West, \& Foster}]{Jones:2011tf}
Jones, D.~O., West, A.~A., \& Foster, J.~B. 2011, AJ, 142, 44

\bibitem[{Kalberla {et~al.}(2005)Kalberla, Burton, Hartmann, Arnal, Bajaja,
  Morras, \& P{\"o}ppel}]{Kalberla05}
Kalberla, P. M.~W. et al. 2005, \aap, 440, 775

\bibitem[{Kashiwagi {et~al.}(2012)Kashiwagi, Yahata, \&
  Suto}]{Kashiwagi:2012we}
Kashiwagi, T., Yahata, K., \& Suto, Y. 2012, ApJS, in press

\bibitem[{McClure-Griffiths {et~al.}(2009)McClure-Griffiths, Pisano,
  Calabretta, Ford, Lockman, Staveley-Smith, Kalberla, Bailin, Dedes,
  Janowiecki, Gibson, Murphy, Nakanishi, \&
  Newton-McGee}]{McClureGriffiths:2009ui}
McClure-Griffiths et al. 2009, ApJS, 181, 398

\bibitem[{M{\'e}nard {et~al.}(2010)M{\'e}nard, Kilbinger, \& Scranton}]{MKS10}
M{\'e}nard, B., Kilbinger, M., \& Scranton, R. 2010, MNRAS, 406, 1815

\bibitem[{Padmanabhan {et~al.}(2008)Padmanabhan, Schlegel, Finkbeiner,
  Barentine, Blanton, Brewington, Gunn, Harvanek, Hogg, Ivezi{\'c}, Johnston,
  Kent, Kleinman, Knapp, Krzesinski, Long, Neilsen, Nitta, Loomis, Lupton,
  Roweis, Snedden, Strauss, \& Tucker}]{Padmanabhan:2008js}
Padmanabhan, N. et al 2008, ApJ, 674, 1217

\bibitem[{Peek \& Graves(2010)}]{2010ApJ...719..415P}
Peek, J. E.~G. \& Graves, G.~J. 2010, ApJ, 719, 415

\bibitem[{Peek {et~al.}(2011{\natexlab{a}})Peek, Heiles, Douglas, Lee,
  Grcevich, Stanimirovi{\'c}, Putman, Korpela, Gibson, Begum, Saul, Robishaw,
  \& Kr{\v c}o}]{2011ApJS..194...20P}
Peek, J. E.~G. et al. 2011{\natexlab{a}}, ApJS, 194, 20

\bibitem[{Peek {et~al.}(2011{\natexlab{b}})Peek, Heiles, Peek, Meyer, \&
  Lauroesch}]{2011ApJ...735..129P}
Peek, J. E.~G.,
  2011{\natexlab{b}}, ApJ, 735, 129

\bibitem[{Peek \& Schiminovich(2013)}]{Peek:2013vg}
Peek, J. E.~G. \& Schiminovich, D. 2013, ApJ, Submitted

\bibitem[{Schlafly \& Finkbeiner(2011)}]{2011ApJ...737..103S}
Schlafly, E.~F. \& Finkbeiner, D.~P. 2011, ApJ, 737, 103

\bibitem[{Schlafly {et~al.}(2010)Schlafly, Finkbeiner, Schlegel, Juri{\'c},
  Ivezi{\'c}, Gibson, Knapp, \& Weaver}]{Schlafly10}
Schlafly, E.~F. et al. 2010, ApJ, 725, 1175


\bibitem[{Schlegel {et~al.}(1998)Schlegel, Finkbeiner, \& Davis}]{SFD98}
Schlegel, D.~J., Finkbeiner, D.~P., \& Davis, M. 1998, ApJ, 500, 525

\bibitem[{Stepnik {et~al.}(2003)Stepnik, Abergel, Bernard, Boulanger, Cambr~sy,
  Giard, Jones, Lagache, Lamarre, Meny, Pajot, Le~Peintre, Ristorcelli, Serra,
  \& Torre}]{Stepnik:2003iq}
Stepnik, B.,et al. 2003, \aap, 398, 551

\bibitem[{Verheijen {et~al.}(2009)Verheijen, Oosterloo, Heald, \& van
  Cappellen}]{Verheijen:2009ua}
Verheijen, M. et al. 2009, in Proceedings of Panoramic Radio Astronomy: Wide-field 1-2 GHz research on galaxy evolution, ed. G. Heald and P. Serra. (Groningen, the Netherlands), http://pos.sissa.it/cgi-bin/reader/conf.cgi?confid=89

\bibitem[{Winkel {et~al.}(2010)Winkel, Kalberla, Kerp, \& Fl{\"o}er}]{Winkel10}
Winkel, B. et al. 2010, ApJS, 188, 488

\bibitem[{Yahata {et~al.}(2007)Yahata, Yonehara, Suto, Turner, Broadhurst, \&
  Finkbeiner}]{Yahata07}
Yahata, K. 2007, PASJ,59, 205

\bibitem[{York {et~al.}(2000)York, Adelman, Anderson, Anderson, Annis, Bahcall,
  Bakken, Barkhouser, Bastian, Berman, Boroski, Bracker, Briegel, Briggs,
  Brinkmann, Brunner, Burles, Carey, Carr, Castander, Chen, Colestock,
  Connolly, Crocker, Csabai, Czarapata, Davis, Doi, Dombeck, Eisenstein,
  Ellman, Elms, Evans, Fan, Federwitz, Fiscelli, Friedman, Frieman, Fukugita,
  Gillespie, Gunn, Gurbani, de~Haas, Haldeman, Harris, Hayes, Heckman,
  Hennessy, Hindsley, Holm, Holmgren, Huang, Hull, Husby, Ichikawa, Ichikawa,
  Ivezi{\'c}, Kent, Kim, Kinney, Klaene, Kleinman, Kleinman, Knapp, Korienek,
  Kron, Kunszt, Lamb, Lee, Leger, Limmongkol, Lindenmeyer, Long, Loomis,
  Loveday, Lucinio, Lupton, MacKinnon, Mannery, Mantsch, Margon, McGehee,
  McKay, Meiksin, Merelli, Monet, Munn, Narayanan, Nash, Neilsen, Neswold,
  Newberg, Nichol, Nicinski, Nonino, Okada, Okamura, Ostriker, Owen, Pauls,
  Peoples, Peterson, Petravick, Pier, Pope, Pordes, Prosapio, Rechenmacher,
  Quinn, Richards, Richmond, Rivetta, Rockosi, Ruthmansdorfer, Sandford,
  Schlegel, Schneider, Sekiguchi, Sergey, Shimasaku, Siegmund, Smee, Smith,
  Snedden, Stone, Stoughton, Strauss, Stubbs, SubbaRao, Szalay, Szapudi,
  Szokoly, Thakar, Tremonti, Tucker, Uomoto, Vanden~Berk, Vogeley, Waddell,
  Wang, Watanabe, Weinberg, Yanny, \& Yasuda}]{york00}
York, D.~G.et al. \aj, 120, 1579

\end{thebibliography}

\end{document}